\RequirePackage{fix-cm}
\documentclass[smallextended]{svjour3}       % onecolumn (second format)
\smartqed  % flush right qed marks, e.g. at end of proof
\usepackage{graphicx}
\begin{document}

\title{Towards a high-precision measurement of the antiproton magnetic moment
%\thanks{Grants or other notes %about the article that should go on the front page should be
%placed here. General acknowledgments should be placed at the end of the article.}
}

%\titlerunning{Short form of title}        % if too long for running head

\author{C.~Smorra$^1$ \and
K.~Blaum$^2$  \and
%S.~Braeuninger \and
K.~Franke$^{1,2}$ \and
%H.~Kracke \and
%C.~Leiteritz \and
Y.~Matsuda$^3$ \and
A.~Mooser$^{4,5}$ \and
H.~Nagahama$^3$ \and
C.~Ospelkaus$^6$ \and
%C.C.~Rodegheri \and
W.~Quint$^7$ \and
G.~Schneider$^{1,4}$ \and
S.~Van Gorp$^8$ \and
J.~Walz$^{4,5}$ \and
Y.~Yamazaki$^8$ \and
S.~Ulmer$^1$ %etc.
}

\authorrunning{C.~Smorra \textit{et al.}} % if too long for running head

\institute{
$^1$ RIKEN, Ulmer Initiative Reseach Unit, 2-1 Hirosawa, Wako, Saitama 351-0198, Japan.
\and
$^2$ Max-Planck-Institut f\"ur Kernphysik, Saupfercheckweg 1, D-69117 Heidelberg, Germany.
\and
$^3$ Graduate School of Arts and Sciences, University of Tokyo, Tokyo 153-8902, Japan.
\and
$^4$ Institut f\"ur Physik, Johannes Gutenberg Universit\"at Mainz, Germany.
\and
$^5$ Helmholtz-Institut Mainz, D-55099 Mainz, Germany.
\and
$^6$ Leibniz Universit\"at Hannover, D-30167 Hannover, Germany.
\and
$^7$ GSI Helmholtzzentrum f\"ur Schwerionenforschung, D-64291 Darmstadt, Germany.
\and
$^8$ RIKEN, Atomic Physics Laboratory, 2-1 Hirosawa, Wako, Saitama 351-0198, Japan.	\newline
\email{christian.smorra@cern.ch}
}

\date{Received: 15.09.2013 / Accepted: date}
% The correct dates will be entered by the editor

\maketitle

\begin{abstract}
The recent observation of single spins flips with a single proton in a Penning trap opens the way to measure the proton magnetic moment with high precision. Based on this success, which has been achieved with our apparatus at the University of Mainz, we demonstrated recently the first application of the so called double Penning-trap method with a single proton. This is a major step towards a measurement of the proton magnetic moment with ppb precision. To apply this method to a single trapped antiproton our collaboration is currently setting up a companion experiment at the antiproton decelerator of CERN. This effort is recognized as the Baryon Antibaryon Symmetry Experiment (BASE). A comparison of both magnetic moment values will provide a stringent test of CPT invariance with baryons.
% \PACS{PACS code1 \and PACS code2 \and more}
% \subclass{MSC code1 \and MSC code2 \and more}
\end{abstract}

\section{Introduction}
\label{intro}
The striking matter-antimatter imbalance observed in our universe is one of the unsolved mysteries of modern physics.
The quantum-field theories involved in the Standard Model (SM) of particle physics, which describe all fundamental interactions except gravity, are CPT invariant. As a consequence, particles and their antiparticles have equal masses, lifetimes, charges and magnetic moments, the latter two with opposite sign. In addition, particles and antiparticles annihilate exactly in collsions. In this context, the matter-antimatter asymmetry lacks a satisfying explanation.
This inspires experiments to compare the properties of matter and antimatter with highest precision (see e.g.~\cite{CPTKmesons,Electron,APM,Muon}); however, no CPT violation has been found so far.\\
One CPT test which has not been carried out with a precision on the ppb level is the comparison of the proton $p$ and antiproton $\overline{p}$ magnetic moments:
\begin{equation}
\mu_{p/\overline{p}} = \frac{g_{p/\overline{p}}}{2}\frac{q_{p/\overline{p}}\hbar}{2m_{p/\overline{p}}},
\end{equation}
where $g$, $q$ and $m$ denote the $g$-factor, the charge and the mass of the particle, respectively. The most precise measurement of $\mu_p$ was performed in 1972 \cite{Winkler1972}, where a relative precision of 10$\,$ppb was achieved. Experiments using single protons in a cryogenic Penning-trap setup are aiming to improve this value by determining the ratio of the spin-precession (or Larmor) frequency $\nu_L$ to the cyclotron frequency $\nu_c$:
\begin{equation}
\frac{\nu_L}{\nu_c} = \frac{g}{2} = \frac{\mu}{\mu_N},
\end{equation}
which directly yields the magnetic moment in units of the nuclear magneton $\mu_N$. 

The cyclotron frequency $\nu_c$ is determined via image-current detection \cite{Wineland} of the motional eigenfrequencies $\nu_+$, $\nu_z$ and $\nu_-$ of the trapped particle by utilizing the invariance theorem \cite{Brownetal1986}. $\nu_L$ is determined by means of the continuous Stern-Gerlach effect \cite{DehmeltCSG}. A strong magnetic field inhomogenity $B_0+B_2z^2$ couples the magnetic moment of the particle to its axial oscillation, where $B_0$ is the magnetic field strength of the Penning trap and $B_2$ characterizes the strength of the magnetic inhomogeneity. The spin state can be determined by observing a frequency shift induced by a spin transition,
\begin{equation}
\Delta\nu_{z,\mathrm{spin}} = \frac{1}{2 \pi^2} \frac{\mu B_2}{m \nu_z}.
\end{equation}
The Larmor frequency $\nu_L$ is extracted by recording the spin-flip probability as function of the frequency $\nu_{rf}$ of an applied spin-flip drive. This scheme has been applied to measure the magnetic moments of the electron and the positron with a precision better than 10$^{-11}$ \cite{RSVanDyck1987}.
However, the (anti)proton magnetic moment is 658 times smaller, which requires an apparatus with much higher sensitivity. For typical axial oscillation frequencies of about 1$\,$MHz, an (anti)proton spin transition shifts the axial frequency by $\Delta\nu_{z,\mathrm{spin}}\approx$ 4$\,$mHz$\cdot B_2/$(T/cm$^2$). Thus, for a successful detection of (anti)proton spin flips a very strong magnetic inhomogeneity is needed. \\
Recently, two independent collaborations reported the first direct measurements of the proton magnetic moment with relative precisions at the ppm level \cite{UlmerPRL2011,Rodegheri2012,JerryProton2012}. These measurements were carried out in a strong magnetic field inhomogeneity of $B_2\approx30\,$T/cm$^2$. The strong $B_2$ broadens the spin transition and the cyclotron line, and thus ultimately limiting the experimental precision to the ppm level. This method was also recently applied to the antiproton and $\mu_{\overline{p}}$ was determined with a relative precision of 4.4$\times 10^{-6}$ \cite{JerryAntiproton2013}.\\
For a further dramatic improvement in experimental precision the so-called double-Penning trap method \cite{Haeffner2003} will be used, which has been applied in magnetic moment measurements of the electron bound to highly charged ions, where relative precisions better than $10^{-10}$ were achieved \cite{Sturm2013}. In this elegant scheme the measurements of $\nu_L$ and $\nu_c$ and the spin-state detection are carried out in two separated traps: a precision trap (PT) with a very homogeneous magnetic field and an analysis trap (AT) with the superimposed magnetic bottle. Recently, we demonstrated the application of the double-trap method for the first time with a single proton \cite{Mooser2013}. This paves a path to a new high-precision measurement of the proton magnetic moment.\\

The Baryon Antibaryon Symmetry Experiment (BASE) aims to apply the double-trap technique to measure the magnetic moment of a single antiproton. The uncertainty of $\mu_{\overline{p}}$ can be improved by at least three orders of magnitude, providing a sensitive test of the CPT invariance. For this purpose, we have constructed a new Penning-trap apparatus, which is based on the apparatus at the university of Mainz used in \cite{UlmerPRL2011,Mooser2013}, but includes significant modifications. BASE will be installed in a recently constructed dedicated new experimental area at the Antiproton Decelerator (AD) of CERN. A new antiproton transfer-line has been designed and gets constructed and installed during the long shutdown (LS1) of CERN to allow operation of BASE in 2014.

\section{The BASE apparatus}
\label{sec:2}

\begin{figure*}
\centering
% Use the relevant command to insert your figure file.
% For example, with the graphicx package use
  \includegraphics[width=0.95\textwidth]{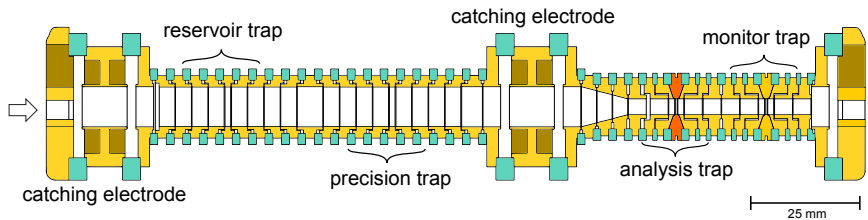}
% figure caption is below the figure
\caption{Schematic of the BASE Penning-trap stack mounted on the pinbase, which serves as electronic interface for the Penning trap chamber. For further details see text.}
\label{fig:1}       % Give a unique label
\end{figure*}

BASE uses a cryogenic Penning-trap system consisting of four traps, as shown in Fig.~\ref{fig:1}. The Penning traps are contained in a pinched-off cryogenic vacuum chamber, which is located in the center of the horizontal bore of a superconducting magnet with a field strength of $B_0$ = 1.9$\,$T. Due to cryopumping, in the trap chamber pressures of about 10$^{-17}\,$mbar can be reached, which will allow antiproton storage times of several months \cite{JerryPbarStorage}.\\
Antiprotons at 5.3 MeV kinetic energy are injected from the AD through a vacuum-tight stainless steel window, which also serves as a degrader \cite{UlmerICPEACProc2013}. Particles with kinetic energies below 15$\,$keV are captured by high-voltage electrodes. Using electron and resistive cooling \cite{JerryPbarCooling}, a cold cloud of antiprotons is prepared in the harmonic potential well of the reservoir trap (RT). Once loaded, this trap serves as an antiproton reservoir, allowing BASE to be operated independently from accelerator cycles and shutdown times. A single antiproton is extracted from the RT into the PT, and will be used for the AT-PT double-trap measurement cycle. The fourth trap in the BASE Penning-trap system is a monitor trap (MT) \cite{PENTATRAP}, which will be used to observe magnetic field fluctuations during the $g$-factor measurement.\\
In the double Penning trap measurement scheme, a single antiproton is prepared and transported to the AT, where its spin-state is determined by a sequence of axial frequency measurements and resonant spin-flip drives. Subsequently, the particle is transported to the PT, where $\nu_c$ is measured via resonant detection of the particle's oscillation frequencies \cite{Ulmer5Dip}, and a spin-flip drive at $\nu_{rf}$ is applied. Afterwards, the particle is transported back to the AT and the spin state is analyzed again. By repeating this measuring sequence for different $\nu_{rf}$, a $g$-factor resonance is obtained \cite{Verdu2004}.\\
This procedure requires single spin-flip resolution \cite{MooserSF2012}, which means that the spin state of the proton has to be clearly identified before and after the PT spin-flip drive. However, this is difficult in the strong magnetic bottle $B_2$, which couples not only the spin magnetic moment to $\nu_z$ but also the angular magnetic moment of the modified cyclotron mode. Fluctuations of the cyclotron energy $E_+$ affect the axial frequency stability, making single spin flips hard to detect. About three quanta $n_+$ in the modified cyclotron mode (3$\times$65 mHz) have a similar effect on the axial frequency as a spin transition (180 mHz). The heating rate of the modified cyclotron mode scales proportional to the modified cyclotron energy $E_+$ \cite{Mooser2013}.
\begin{figure*}
\centering
% Use the relevant command to insert your figure file.
% For example, with the graphicx package use
  \includegraphics[width=0.75\textwidth]{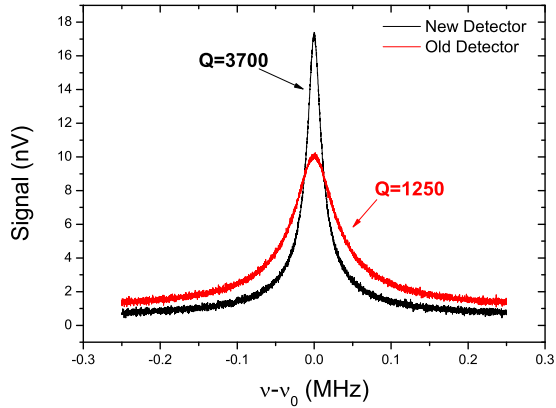}
% figure caption is below the figure
\caption{Resonances of the cyclotron detection systems. Black: New detection system which will be used in BASE (Q=3700), and red: detection system used in \cite{Mooser2013} with Q=1250. The quality factor of the new detection system is about a factor of 3 improved.}
\label{fig:2}       % Give a unique label
\end{figure*}
At a threshold of $E_{+,th} < $ 150 $\mu$eV, we achieve a detection fidelity of 75$\,\%$, which means that in three out of four spin measurements the eigenstate is identified correctly. To prepare a cyclotron state with $E_+<E_{+,th}$ resistive cooling is used. 

The particle is cooled by the effective resistance $R_p$ of the detection system, which is based on an inductor $L$ followed by a cryogenic ultra low-noise amplifier connected to a trap electrode. Together with the trap capacitance $C_T$ it forms a tuned circuit with resonance frequency $\nu_0 = (2 \pi \sqrt{L C_T})^{-1}$. On resonance it acts as an effective parallel resistance $R_p = 2\pi\nu_0 Q L$, where $Q$ is the quality factor of the tuned circuit. Image currents induced by the oscillating particle in the trap electrodes dissipate energy in the detection system \cite{Wineland} and the particle is cooled with a cooling time constant
\begin{equation}
\tau = \frac{m D^2}{q^2 R_p},
\end{equation}
where $D$ is a trap specific length. Preparing the particle after a measurement of $\nu_c$ for a spin-state read out below the threshold energy $E_{th}$ required on average about two hours with the detector used in \cite{Mooser2013}. Thus, we developed for BASE a significantly improved cyclotron detection system. As inductor, a superconducting NbTi coil is placed inside a cylindrical copper housing with a diameter of about 25$\,$mm. The coil has an inductance of $L$ = 2.1 $\mu$H and an unloaded quality factor ($Q$ value) of 11400. When connected to the ultra low-noise amplifier and the trap, a $Q$ value of 3700 is achieved. A fast Fourier spectrum of the noise resonance of the new detection system is shown in Fig.~2, black line. Compared to the previously used detection system (Fig.~2, red line) the quality factor is improved by about a factor of 3. This will enable BASE to perform much faster double-trap cycles as reported in \cite{Mooser2013}.\\
\begin{figure*}
\centering
% Use the relevant command to insert your figure file.
% For example, with the graphicx package use
  \includegraphics[width=0.75\textwidth]{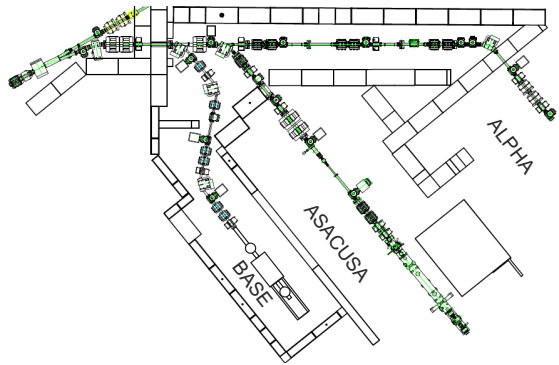}
% figure caption is below the figure
\caption{Layout of the AD ejection beamlines, including the new beamline for BASE and its experimental area. }
\label{fig:3}       % Give a unique label
\end{figure*}
In April 2013, the BASE project has been accepted by the CERN research board. The experiment will be implemented into the AD facility until the end of the accelerator shutdown LS1 in 2014. Fig.~\ref{fig:3} shows a top view of the BASE zone. The new antiproton transfer-line has been designed and verified by the AD group of CERN and is currently under construction.\\

In conclusion, the Baryon Antibaryon Symmetry Experiment BASE aims for a measurement of the antiproton magnetic moment with a relative precision of at least 10$^{-9}$ by applying the double Penning-trap method. This will provide a sensitive test of CPT invariance with Baryons. Installation work for the BASE experimental setup in the AD are ongoing.

\begin{acknowledgements}
We would like to express our gratitude towards L.~Bojtar, T.~Eriksson, F.~Butin, T.J.~Rutter and S.~Maury, who are responsible for the implementation of the BASE into the AD infrastructure. We are also thankful towards all CERN groups which contribute to BASE. We acknowledge financial support of RIKEN Initiative Research Unit Program, RIKEN president fund, the Max-Planck Society, the BMBF, the Helmholtz-Gemeinschaft, and the EU (ERC Grant No. 290870-MEFUCO).
\end{acknowledgements}

% BibTeX users please use one of
%\bibliographystyle{spbasic}      % basic style, author-year citations
%\bibliographystyle{spmpsci}      % mathematics and physical sciences
%\bibliographystyle{spphys}       % APS-like style for physics
%\bibliography{}   % name your BibTeX data base

% Non-BibTeX users please use

\end{document}